\documentclass[conference]{IEEEtran}
\usepackage{epsfig}
\usepackage{subfigure}
\usepackage{calc}
\usepackage{amstext}
\usepackage{amsmath}
\usepackage{amsthm}
\usepackage{multicol}
\usepackage{pslatex}

\usepackage{amssymb}
\usepackage{graphicx}
\usepackage{listings}
\usepackage{color}
\usepackage{subfig}
\usepackage{multirow}

 \usepackage{pbox}
\usepackage[usenames,dvipsnames,svgnames,table]{xcolor}
\usepackage{url}
\definecolor{gray}{rgb}{0.4,0.4,0.4}
\definecolor{darkblue}{rgb}{0.0,0.0,0.6}
\definecolor{cyan}{rgb}{0.0,0.6,0.6}

\lstdefinelanguage{XML}
{
  morestring=[b]",
  morestring=[s]{>}{<},
  morecomment=[s]{<?}{?>},
  stringstyle=\color{black},
  identifierstyle=\color{darkblue},
  keywordstyle=\color{cyan},
  morekeywords={xmlns,version,type}
}

\begin{document}
%
\title{Towards Industry 4.0: Gap Analysis between Current Automotive MES and Industry Standards using Model-Based Requirement Engineering}

\author{\IEEEauthorblockN{
S. Manoj Kannan\IEEEauthorrefmark{1}\IEEEauthorrefmark{2},
Kunal Suri\IEEEauthorrefmark{1}\IEEEauthorrefmark{4},
Juan Cadavid\IEEEauthorrefmark{1},
Ion Barosan\IEEEauthorrefmark{2},
Mark van den Brand\IEEEauthorrefmark{2},\\
Mauricio Alferez\IEEEauthorrefmark{1}, and
Sebastien Gerard\IEEEauthorrefmark{1}} 
\IEEEauthorblockA{\IEEEauthorrefmark{1}CEA, LIST, Laboratory of Model Driven Engineering for Embedded Systems (LISE), P.C. 174, \\ Gif-sur-Yvette, 91191, France \\
}
\IEEEauthorblockA{\IEEEauthorrefmark{2}Eindhoven University of Technology (TU/e),
Eindhoven, The Netherlands\\
}
\IEEEauthorblockA{\IEEEauthorrefmark{4}Telecom SudParis, UMR 5157 Samovar, Universit\'e Paris-Saclay, France \\ 
\{manojkannan.soundarapandian, kunal.suri, juan.cadavid, mauricio.alferez, sebastien.gerard\}@cea.fr, \\ \{i.barosan, m.g.j.v.d.brand\}@tue.nl
}}


\maketitle

\begin{abstract}
The dawn of the fourth industrial revolution, Industry 4.0 has created great enthusiasm among companies and researchers by giving them an opportunity to pave the path towards the vision of a connected smart factory ecosystem. However, in context of automotive industry there is an evident gap between the requirements supported by the current automotive manufacturing execution systems (MES) and the requirements proposed by industrial standards from the International Society of Automation (ISA) such as, ISA-95, ISA-88 over which the Industry 4.0 is being built on. In this paper, we bridge this gap by following a model-based requirements engineering approach along with a gap analysis process. Our work is mainly divided into three phases, (i) automotive MES tool selection phase, (ii) requirements modeling phase, (iii) and gap analysis phase based on the modeled requirements. During the MES tool selection phase, we used known reliable sources such as, MES product survey reports, white papers that provide in-depth and comprehensive information about various comparison criteria and tool vendors list for the current MES landscape. During the requirement modeling phase, we specified requirements derived from the needs of ISA-95 and ISA-88 industrial standards using the general purpose Systems Modeling Language (SysML). During the gap analysis phase, we find the misalignment between standard requirements and the compliance of the existing software tools to those standards.

\end{abstract}


%
\IEEEpeerreviewmaketitle

\section{Introduction}
\label{sec:introduction}

The manufacturing sector is considered as the backbone for facilitating economic growth of a country~\cite{westkamper2014manufacturing}. Today, it is difficult to imagine managing the production operations in an industry without the use of automation, computer systems and software. However, the ever-increasing complexity of the software-intensive systems being used in automotive manufacturing industry has created the need to provide processes having simple and human understandable tools along with creation of modular and adaptable systems. Moreover, to maintain competitiveness in the automotive manufacturing sector the factories of today have to mature into something smarter and efficient, e.g., by reducing wastage of materials and energy.

The fascination for Industry 4.0 can be described from two point of views, first, this is one of the first kind of industrial revolution which is predicted a-priori, not observed ex-post. This provides opportunities to companies and research institution to actively shape the future of manufacturing sector. Secondly, Industry 4.0 promises a huge economic impact as it envisions a substantial increase in operational effectiveness along with the development of entirely new business models, services and products \cite{hermann2016design}. In other words, Industry 4.0 is the next industrial revolution which is about to take place now. Not to forget, there are several parallel initiatives taking place globally such as the \textit{Factories of the Future} (FoF) along with institutions such as the \textit{National Institute of Standards and Technology} (NIST) \cite{lu2016current} actively involved in proposing standards for the upcoming fourth industrial revolution.

One important thing to note about the Industry 4.0 vision is that there are much work describing what Industry 4.0 could do, why it is important and what standards could be used. However, to the best of our knowledge there are not much work available in literature that describe steps to be taken by companies who wish to move towards the Industry 4.0 based on the current industrial standards and requirements. Thus, in this paper we provide an approach to suggest the most important requirements that any ``automotive'' manufacturing execution systems (MES) tool must support in order to comply with current industrial standards from the International Society of Automation (ISA) such as, ISA-95, ISA-88 and therefore, be prepared for upcoming standards such as the {\em Reference Architecture Model for Industry 4.0} (RAMI 4.0). 

Likewise, the objectives of our research presented in this paper are as follows: \textit{objective-1}: to define the criteria used for selecting MES vendors and their tools (automotive domain), \textit{objective-2}: to understand the current tool landscape and their compliance with the standards mainly ISA-95 (building block for RAMI 4.0) but also ISA-88, and \textit{objective-3}: to analyze the gap between the existing automotive MES tools and ISA-95.

The rest of the paper is structured as follows: in Section \ref{sec:background}, we briefly introduced some concepts required to understand our work in a better way. In Section \ref{sec:approach}, we present our approach followed by a brief discussion in Section \ref{sec:discussion}. In Section \ref{sec:conclusion}, we conclude our paper and talk about the future work.
\section{Background}
\label{sec:background}
In this Section, we briefly introduce model-based requirement engineering in Section \ref{MBSE_RE}. In Section \ref{isa95_standard}, we provide a brief information about ISA-95 standard while in Section \ref{rami40}, we introduce the Reference architecture model for Industry 4.0 (RAMI 4.0).

\subsection{Model-Based Requirement Engineering}
\label{MBSE_RE}
Model-driven engineering (MDE) is the systematic use of models as primary artifacts during a software engineering process~\cite{hutchinson2014model} for handling software complexity. MDE improves productivity gain, portability, maintainability, understanding and separation of concerns thus helps in reducing system complexity. By separating the control flow and execution flow, flexibility of the models automatically increases. Another important benefit of MDE is simulation and visualization of models. This fosters confidence on the behavior of the system before implementing it in the real world. 

Furthermore, model-based techniques in requirement engineering~\cite{alferez2008model,Rashid:2003:MCA:643603.643605,bombonatti2016usability} help to understand the relationship between various requirements based on standard languages such as SysML for requirements. A requirements model fosters understanding in following ways:
\begin{itemize}
\item Describe the users, customers and other stakeholders needs with less ambiguity
\item Reduces inconsistencies and support tractability of requirements
\item Uses models as a basis for system testing by explicitly defining the relationships between test cases and requirements
\item Reduces the efforts required to make changes in the requirement during evolution. Due to predefined relations, any requirement evolution will update the test accordingly. This will ensure that the system always meets the new or updated requirements
\end{itemize}

\subsection{The ISA-95 Standard}
\label{isa95_standard}
To have a standardized development of information systems for all manufacturing industries the International Society of Automation (ISA) proposed ISA-95\footnote{\url{https://www.isa.org/isa95/}}~\cite{isa95,isa954}, which provide definitions about the MES functions and the data exchanged between the enterprise resource planning system (ERP) and control systems (MES or other control systems). The ISA-95 standard describes the interface between the manufacturing operations and control functions along with other enterprise functions. One of the important interfaces defined in ISA-95 is the one between the level 3 and level 4 of the functional hierarchical model. This interface helps to integrate the enterprise and control systems so that they work together for producing any manufacturing product effectively.

\subsection{Reference Architecture Model for Industry 4.0 (RAMI 4.0)} 
\label{rami40}
\begin{figure}[ht]
\centering
\includegraphics[width=0.48\textwidth]{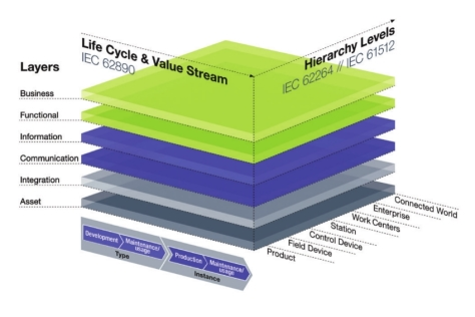}
\caption{Reference Architecture Model for Industry 4.0 (Source:ZVEI)}
\label{fig:rami_40}
\end{figure}

RAMI 4.0 \cite{hankel2015reference} provides a common understanding of the relations existing between various individual components for Industry 4.0 solutions landscape. It also provides a common viewpoint for different industry branches. As illustrated in Figure \ref{fig:rami_40} (source:ZVEI\footnote{ZVEI:\url{http://www.zvei.org/en/subjects/Industry-40/Pages/The-Reference-Architectural-Model-RAMI-40-and-the-Industrie-40-Component.aspx}}), RAMI 4.0 is composed of a three dimensional coordinate system which describes all the important aspects of Industry 4.0. The three axes makes it possible to map all critical aspects of Industry 4.0 making it a kind of 3D mapping of Industry 4.0 solutions. It provides a direction for plotting the requirements of various field together with different standards in order to implement Industry 4.0 techniques and enables identification of overlapping requirements, gaps and their resolution based on the related industrial standards.

    \section{Approach}
\label{sec:approach}

In this Section, we detail the approach followed in our work as follows: In Section \ref{Vendor_Selection_Phase}, we perform the vendor selection for MES tools from the automotive domain. In Section \ref{sec:Requirement_Modeling}, we model the requirements using SysML. In Section \ref{sec:gap_analysis}, we perform the gap analysis to find the misalignment between the standard requirements and the compliance of the existing MES tools to those standards. Finally, in Section \ref{sec:approach_recapitulation}, we provide a recapitulation of our approach.

\subsection{Vendor Selection Phase}
\label{Vendor_Selection_Phase}
In this paper, we keep our focus on the automotive domain, mainly to understand the gap between the existing MES tools and the industrial standards. Moreover, the vendor selection process starts with the extraction of information from highly reliable sources such as, the sixteenth edition (September 2015) of the annual MES product survey report from CGI Group \cite{ CGI_Group_MES_Product_Survey}. This report provides exhaustive details of 71 MES products for 67 vendors. Apart from this survey report, we analyzed white papers, online blogs and publications \cite{Business_Software_advice_report, Capterra_advice_report,software_advice_report,plex_manufacturing,CTS_guide_manuf} based on the popularity of these sources (number of users visitors) along with the duration they have been involved in providing information about MES tools. On the bases of the above mentioned sources, we shortlisted the most common vendors as shown in Table \ref{Shortlisted Automotive MES Vendors and their product}.

\begin{table}[ht]
\centering
\resizebox{0.49\textwidth}{!}{
\begin{tabular}{cll}
\\ \hline \\
\textit{\textbf{S.No}} & \textit{\textbf{Vendor Name}} & \textit{\textbf{Product Name}} 
\\ \hline \\
1                      & ABB                           & cpmPlus Suite                  \\
2                      & Dassault Systems              & DELMIA Apriso                  \\
3                      & Schneider Electric            & Ampla                          \\
4                      & Siemens                       & SIMATIC IT                    \\
5                      & Honeywell Process Solutions   & Intuition                      \\
6                      & Rockwell Automation           & FactoryTalk                    \\
7                      & SAP                           & SAP MII                        \\
8                      & Oracle                        & Oracle MES                     \\
9                      & GE Intelligent Platforms      & Proficy for Manufacturing     
\\ \hline
\end{tabular}}
\caption{Shortlisted Automotive MES Vendors and their product}
\label{Shortlisted Automotive MES Vendors and their product}
\end{table}

After the vendor selection, we evaluate the MES tools based on certain quality criteria. These criteria for MES evaluation must cover both, the product technical details and the vendor portfolio. For doing this, we created six main categories, they are, \textit{(i) Corporate vision, (ii) Technology and system architecture, (iii) Product functionality, (iv) Product-Industry target, (v) Service and support, and (vi) Supplier longevity}. These categories have been created in such a way that any technical or non-technical feature of a MES tool must fit in one of them. After this step, a list of criteria are formulated and a priority level is assigned for each criterion. Due to limitation of space, we provide only one example of the above mentioned criteria without going into details for the rest of them. 

The example is as follows: \textbf{Technology and System Architecture}\\
\textit{Criteria} : \textbf{Architecture Compatibility}\\
\textit{Priority Level} : 3 \\
\textit{Conditions} : 
\begin{itemize}
\item Score 1 - Compatible with less than 3 architectures
\item Score 2 - Compatible between 3 and 5 architectures
\item Score 3 - Compatible with more than 5 architectures
\end{itemize}
\textit{Comments} : This criteria shows the compatibility of products while applying some architectures as basis (parts of) for the applications. Since an architecture act as a framework of product in its lifetime, top priority level of 3 has been assigned. In the current trend, software tools supports more number of architectures to have better compatibility during implementation.  The above mentioned conditions have been chosen based on the trends observed from the CGI survey report \cite{ CGI_Group_MES_Product_Survey}

After each vendor has been evaluated against the various criteria, an overall collective score is calculated as shown in Table \ref{Vendor Selection Process}

\begin{table}[h]
\centering
\resizebox{0.49\textwidth}{!}{
\begin{tabular}{cm{4cm}ccm{1.5cm}}
\\ \hline
\textit{\textbf{S.No}} & \textit{\textbf{Vendor Name}} & \textit{\textbf{Score}} & \multicolumn{1}{l}{\textit{\textbf{\begin{tabular}[c]{@{}l@{}}ISA 95 Compliance\\       (percentage)\end{tabular}}}} & \textit{\textbf{\begin{tabular}[c]{@{}c@{}}Remarks\\ (Specific\\ /not Specific) \\ for Automobile \\ Industry\end{tabular}}} 
\\ \hline \\
1                      & Rockwell Automation           & 99                      & 91.5                                                                                                                 & Specific                                                                                                                     \\
2                      & SAP                           & 97                      & 18.25                                                                                                                & Not Specific                                                                                                                 \\
3                      & ABB                           & 97                      & 93.25                                                                                                                & Not Specific                                                                                                                 \\
4                      & GE Intelligent Platforms      & 96                      & 96.25                                                                                                                & Specific                                                                                                                     \\
5                      & Dassault Systems              & 96                      & 100                                                                                                                  & Specific                                                                                                                     \\
6                      & Siemens A\&D AS MES           & 88                      & 85.5                                                                                                                 & Not Specific                                                                                                                 \\
7                      & Oracle                        & 80                      & 92.5                                                                                                                 & Not Specific                                                                                                                 \\
8                      & Honeywell Process Solutions   & 78                      & 56                                                                                                                   & Not Specific                                                                                                                 \\
9                      & Schneider Electric            & 60                      & 56.5                                                                                                                 & Not Specific                                                                                                                
\\ \hline
\end{tabular}}
\caption{Vendor Selection Process}
\label{Vendor Selection Process}
\end{table}

From Table \ref{Vendor Selection Process}, illustrates that most of the MES software product that have a maximum score also comply to the ISA-95 standard. This shows that customer requirements and manufacturing standards are well aligned and MES software developing vendors are also fulfilling these requirements. 

However, the {\em SAP MII} tool is an exception as it does not satisfy the ISA-95 standard (with only 18\% compliance). This is mainly because they make most of the features as \textit{custom based} and do not provide out-of-the-box and configurable options. Based on this detailed analysis, automotive MES products with top three scores are selected for benchmarking purpose and for framing the specification that could be used to create an state of the art MES tool. Plus, in order to have a comparative study between automotive specific and non-automotive specific MES product, ABB product is also taken into consideration (even though it is not specific to automotive application) as it has the highest level of compliance with ISA-95. The selected MES products are shown in Table \ref{Selected MES Products for Benchmarking Study}.

\begin{table}[h]
\centering
\resizebox{0.49\textwidth}{!}
{\begin{tabular}{cll}
\hline \\
\textit{\textbf{S.No}} & \textit{\textbf{Vendor Name}} & \textit{\textbf{MES Software Name}}
\\ \hline \\
1                      & Rockwell Automation           & FactoryTalk                         \\
2                      & ABB                           & cpmPlus Suite                       \\
3                      & GE Intelligent Platforms      & Proficy for Manufacturing           \\
4                      & Dassault Systems              & DELMIA Apriso                      
\\ \hline 
\end{tabular}}
\caption{Selected MES Products for Benchmarking Study}
\label{Selected MES Products for Benchmarking Study}
\end{table}

\subsubsection{Product Analysis}
The four top shortlisted MES tools are used to benchmark the product features for a state of the art MES tool development. Though each tool has its own unique features for satisfying some specific customer requirements, there are some common features that most of the software tools provides in their packages. All these features are reviewed and recommended \cite{CGI_Group_MES_Product_Survey} for development of MES:
\begin{itemize}
\item \textbf{Platforms}: The supported platforms (operating systems) have been split into two groups: server and client systems
\item \textbf{Database}: Oracle and SQL Server are dominant at the supported database systems
\item \textbf{Technology}: It is recommended to support technologies such as Com-technology, MS-platform, JAVA platform, ODBC and HTML5 for MES product development
\item \textbf{Architecture}: It is recommended to apply Client/Server, distributed and SOA architecture as all of the benchmarking tools supports the same
\item \textbf{Validation and Regulatory Compliance}: It is recommended to comply to GAMP regulated environment, 21 CFR part 11 and validation dossier MES development
\item \textbf{Languages}: It is recommended to support \textit{Full-NLS}, Full National Language Support for improved user comfortness and Unicode standard
\item \textbf{Incident Management}: It is recommended to support \textit{Detecting latent incidents} and have both \textit{Activate tracing} and \textit{De-activate tracing}
\end{itemize}

\subsection{Requirement Modeling Phase}
\label{sec:Requirement_Modeling}
In our approach, we used the ISA-95 industrial standard as a reference to collect the needs for developing a MES tool. The ISA-95 is one of the building blocks for RAMI 4.0. Additionally, we also cross checked with ISA-88 standard (as it is very close to ISA-95). Once we collected all the needs from ISA-95, we framed the requirements from the needs and modeled them using SysML. We found a total of \textbf{106} requirements and we grouped them by functionality given in the ISA-95 standard~\cite{isa95}. This grouping of requirements is reflected in the requirements model created using SysML in form of packages. 

\begin{figure}[ht]
\centering
\includegraphics[width=0.49\textwidth]{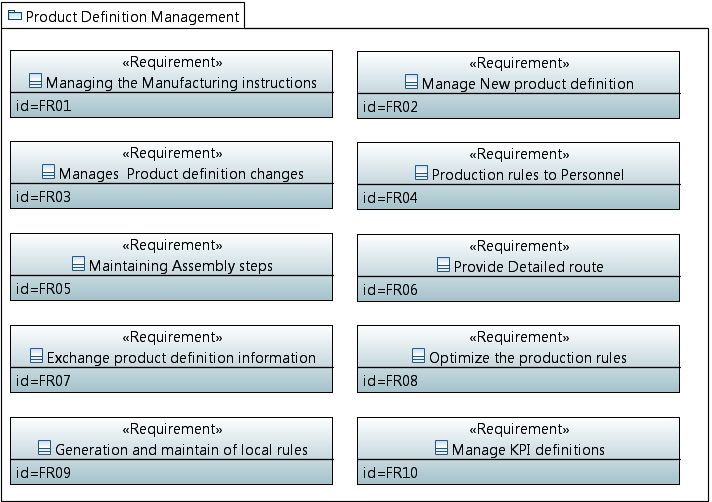}
\caption{Product Definition Management Package}
\label{fig:Product-Definition-Management-Package}
\end{figure}

Due to space limitations, we provide example of one of the Package containing various requirements. Figure \ref{fig:Product-Definition-Management-Package} depicts the \textit{Product Definition Management Requirements Package} and Figure \ref{fig:Product-Definition-Management-Requirements-Model} depicts the \textit{Product Definition Management Requirements Model} based on ISA-95.

Figure \ref{fig:Product-Definition-Management-Package}, illustrates the 10 available requirements that are present in this package, they are:
 \begin{itemize}
 \item FR01 - \textbf{Managing the Manufacturing instructions} - The software shall manage the work masters, manufacturing instructions, recipes, product structure diagrams, manufacturing bills and product variant definitions
 \item FR02 - \textbf{Manage New Product Definition} - The software shall manage the new product definitions
 \item FR03 - \textbf{Manages Product Definition Changes} - The software shall manage changes to product definition
 \item FR04 - \textbf{Production Rules to Personnel} - The software shall provide the product production rules to personnel and relevant activities
 \item FR05 - \textbf{Maintaining Assembly Steps} - The software shall maintain the feasible detailed production routing for products
 \item FR06 - \textbf{Provide Detailed Route} - The software shall provide the detailed route to the manufacturing operations
 \item FR07 - \textbf{Exchange Product Definition Information} - The software shall manage the exchange of product definition information with Level 4 as per the business operation requirement
 \item FR08 - \textbf{Optimize the Production Rules} - The software shall optimize the product production rules based on process analysis and production performance analysis
 \item FR09 - \textbf{Generation and Maintenance of Local Rules} - The software shall generate and maintain the local production rule sets related to the product for cleaning, start-up and shutdown operations
 \item FR10 - \textbf{Manage KPI Definitions} - The software shall manage the key performance indicator (KPI) definitions associated with the products and production
 \end{itemize}
 
 Apart from the above mentioned requirements, there are two more requirements involved in the \textit{Product Definition Management Requirements} model as illustrated in the Figure \ref{fig:Product-Definition-Management-Requirements-Model}. They are:
\begin{itemize}
    \item FR60 - \textbf{Generate Production Performance Information} - The software shall generate production responses and production performance information.
    
    \item FR75 - \textbf{Certify the Product Quality} - The software shall perform product quality certification.
\end{itemize}

\begin{figure}[ht]
	\begin{center}
	\includegraphics[width=0.27\textwidth]{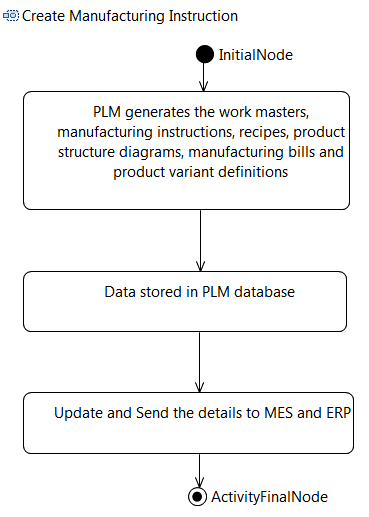}
	\caption{Activity Diagram for creating Manufacturing Instruction}
	\label{fig:Activity_diagram_for_Manufacturing_instruction_process}
	\end{center}
\end{figure}

\textbf{Requirements model and relationships}: In the requirements model Figure \ref{fig:Product-Definition-Management-Requirements-Model}, the relationship are identified and requirements are grouped into categories. The relationship identified in this package are explained as follows:
\begin{itemize}
\item The requirement ID FR60 and FR75 are derived from FR10. A derive relationship may either refine or restate a requirement to improve stakeholder communications or to track design evolution. In this case, FR60 deals with generation of production performance information and FR75 deals with product quality certification. These requirements restate the KPI of product and production together which is specified in FR10. Both derived requirements can survive on its own without FR10 support. It only shares the information. Hence \textit{derive relation} is preferred than \textit{composition}.

\begin{figure}[ht]
\centering
\includegraphics[width=0.48\textwidth]{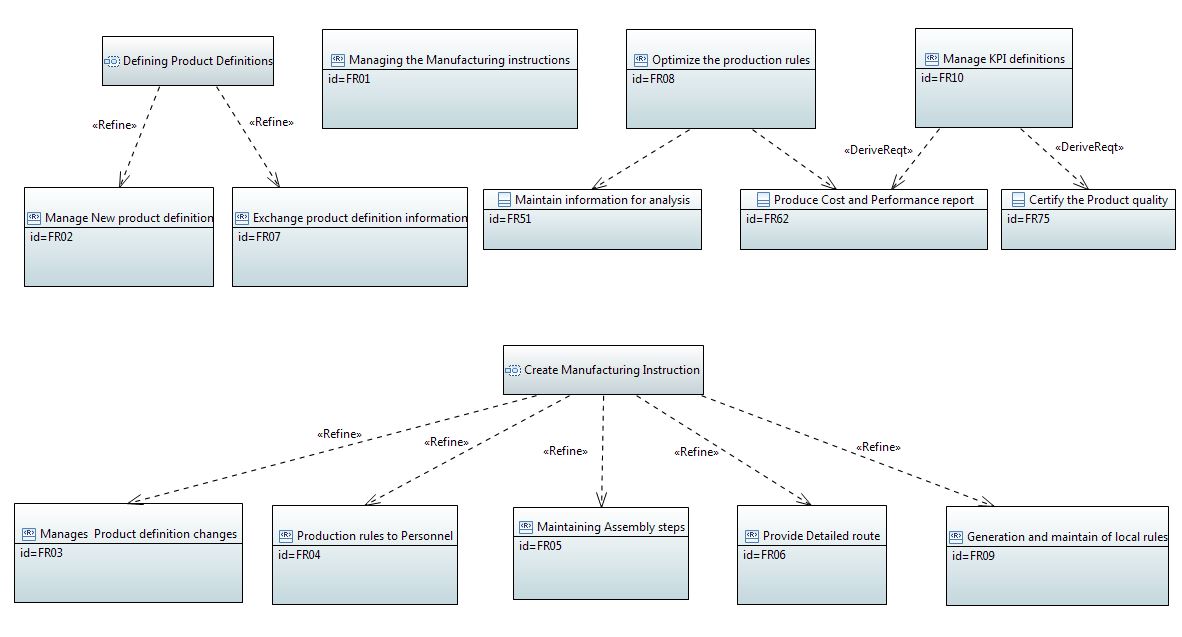}
\caption{Product Definition Management Requirements Model}
\label{fig:Product-Definition-Management-Requirements-Model}
\end{figure}

\item Requirements are refined and realized by use case scenarios. In this case, FR02 and FR07 are refined by activity model \textit{Defining product definitions} shown in Figure \ref{fig:Activity_diagram_for_Product_Definition}. This model explains the flow of information from PLM process to MES and ERP systems. One of the key focus of ISA-95 standard is the integration of information between ERP and other hierarchical level.

 \begin{figure}[ht]
	\begin{center}
	\includegraphics[width=0.27\textwidth]{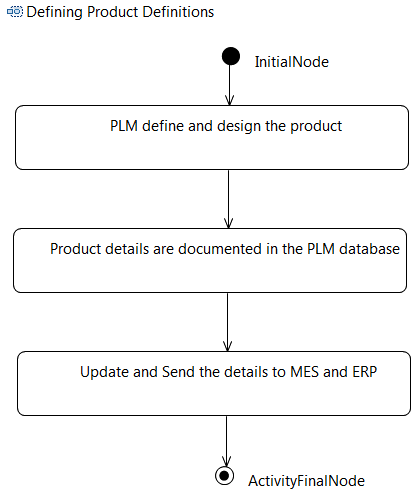}
	\caption {Activity Diagram for Product Definition}
	\label{fig:Activity_diagram_for_Product_Definition}
	\end{center}
\end{figure}

\item FR03, FR04, FR05, FR06 and FR09 are refined by activity model \textit{Creating manufacturing instructions} shown in Figure \ref{fig:Activity_diagram_for_Manufacturing_instruction_process}. This model explains the various processes involved in PLM. The output of these processes acts as an input data for lower levels in the functional hierarchy. Only top level data flow is shown in this model since PLM is quite big topic to be analyzed. To explain PLM in nutshell, this system gives the life for any product since it deals from initial design till it reaches the customers. For an ideal integration, all information should flow from PLM to other manufacturing information systems.
\item FR01 is modelled in the production dispatching package since few requirements are derived from it.
\end{itemize}

Overall during the modeling phase, each requirement is studied and relationship between requirements is analyzed to develop the correct requirements model. Initially from manufacturing standard (i.e. ISA 88 and ISA 95), \textbf {106} requirements are extracted. After modeling them, it is identified that almost all requirements are interlinked and share information between them. Detailed analysis of this requirements models shows that there are \textbf{33} functional requirements that acts as the root requirements. Remaining requirements are related to these root requirements either by composition, derivative relation or abstraction.

\subsection{Gap Analysis Phase}
\label{sec:gap_analysis}
In this Section, we detail the steps involved in the gap analysis phase that we followed in our work.

\subsubsection{Product - Standard Compliance Details}
\label{Product_Standard_Compliance_Details}
Before starting the gap analysis, it is necessary to know how Automotive MES products fulfills the requirements of ISA-95 standard. This helps to pin-point the precise area of improvements and provides a way to know how the existing software performs. To get inputs for this process, technical reports and literature surveys acts as the main source of information \cite{CGI}. MES functionalities are divided into four categories as per the ISA-95 standard that are, (i) \textit{Production Operations Management}, (ii) \textit{Maintenance Operations Management}, (iii) \textit{Quality Operations Management}, and (iv) \textit{Inventory Operations Management}. Using the data from the reports, compliance percentage of MES products w.r.t standards for each category are found \cite{CGI}.

Based on the MES product compliance (for ISA-95) from the survey report \cite{CGI} we checked 37 automobile MES products and found that only two of them comply with all the requirements of manufacturing standards. They are \textit{DELMIA Apriso} and \textit{DIAMES}. However, there were 12 automobile MES products that comply with more than 90\% of the requirements. Plus, in total 19 tools were found to have a compliance of 75\% and above. 

\subsubsection{Steps Involved in Gap Analysis}
For the gap analysis, a list of 37 vendors were selected. To point back, during the vendor selection phase (see Section \ref{Vendor_Selection_Phase}), we had selected 4 vendors for benchmarking the common MES features but in gap analysis phase we use a list of 19 vendors who have more than 75\% compliance. Furthermore, the 33 root requirements found during requirements modeling are listed against each of these 19 vendors. Using the vendor compliance data from the survey, a scale range of 1 to 5 would be assigned to each of the 33 requirement for each of 19 MES vendors. The weightage of the score is are explained below :
\begin{itemize}
\item Score 1 - None of the features that satisfy the requirements are available in the software tool
\item Score 2 - Features are available but they are completely custom made features that are developed and installed based on customer request
\item Score 3 - Features are available but in the form of libraries. Therefore, only if these libraries are installed, requirements get satisfied.
\item Score 4 - Features are available to satisfy the requirements in the software tool but modifications in the configuration are required
\item Score 5 - Features are completely out-of-the-box and no modifications are required to satisfy the requirements. They fulfill all the standard requirements
\end{itemize}

Since the survey reports uses ISA-95 as a reference standard to get the vendor feedback about their products, it is perfectly aligned with our work. Once the scores have been assigned to each of the 33 requirement against each of the 19 MES product, overall scores for each requirement is calculated. Thus, the maximum score for each requirement will come out to be 95 (19 X 5). From the overall scores, we identified the priority requirements.

\subsubsection{Gap Analysis Result}
\label{sec:gap_analysis_result}
For the gap analysis, out of a total score of 95 for each requirement, we segregated and prioritized requirements that had a score of 85 and above. Ine general, requirements with a score of 85 or above were satisfied by more than 75\% of the MES products. Thus, if a MES tool satisfies these prioritized requirements, it will help in solving the challenges faced by the automotive industries. These prioritized requirements are presented as follows (highest to lowest priority):
\begin{enumerate}
\item The software shall perform product quality certification
\item The software shall determine the bottleneck resources for each period and ensure the time of future production availability for particular production
\item The software shall ensure that the equipment is available for the assigned tasks and that job titles are correct
\item The software shall report on inventory to production, quality, maintenance operations management and/or Level 4 activities
\item The software shall measure and report on inventory and material transfer capabilities
\item The software shall reserves the resources for future use
\item The software shall manage the exchange of product definition information with Level 4 as per the business operation requirements
\item The software shall inform the detailed production scheduling when the scheduled requirements are not met
\item The software shall provide information on resource capability
\item The software shall support supervising the requested maintenance
\item The software shall make performance verification of production equipment's
\item The software shall generate and maintain the local production rule sets related to products for cleaning, start-up and shutdown operations
\end{enumerate}

In other words, the above 12 requirements act as a starting point for the development of a MES tool. Additionally, for selecting the functionalities that are needed to be develop in form of a requirement model, there is a need to observe the relationships between each these requirements. From our observations, we found that majority of the requirements are satisfied if the requirement models are developed for the \textit{Performance Analysis} functionality. This further narrows down the proposal for model development to \textit{Performance Analysis}. Additionally, out of these 12 requirements, nearly half belongs to Quality, Maintenance and Inventory Operations management all-together (see Section \ref{Product_Standard_Compliance_Details}). The remaining half also indirectly depend on the performance of these functionalities. Therefore, there is an evident need of improvements w.r.t Quality, Maintenance and Inventory management and not only concerning the production operation management.

\subsection{Approach Recapitulation}
\label{sec:approach_recapitulation}
During the \textit{Vendor Selection Phase} (see Section \ref{Vendor_Selection_Phase}), we selected top four MES providers based on their compliance with ISA-95 standards (three automotive and one generic (having 100\% compliance)). These MES providers scored high on both our quality criteria and the compliance level with ISA-95, as illustrated in Table \ref{Vendor Selection Process}.

During the \textit{Requirement Modeling Phase} (see Section \ref{sec:Requirement_Modeling}), we modeled the requirements based on the needs defined in ISA-95 and ISA-88. Moreover, in this step, we extracted a total of 106 requirements from our interpretation of the needs provided in the standards. After extracting these needs and modeling them using SysML, we modeled the dependencies between these requirements. During modeling, we found out that there are 33 root requirements on which all other requirements depend on. 

During the \textit{Gap Analysis Phase} (see Section \ref{sec:gap_analysis}), we used a list of 37 automotive MES vendors based on the MES product survey report~\cite{CGI}. We found that only two automotive MES tools comply 100 \% with the ISA-95 standard, while 19 MES tools had 75\% or more compliance rate. Thus, to do the gap analysis, we took these 19 MES tools and mapped each of them against the 33 root requirements derived from \textit{Requirement Modeling Phase}. Then, we segregated the requirements that had an overall score of 85 or more (out of 95: 19 tools X maximum score of 5). In total, we found that there are 12 root/parent requirements which satisfy the major portion of the gap identified and act as a starting point to build a state of the art MES tool for automotive domain.

\section{Discussion}
\label{sec:discussion}

This work is motivated towards assisting the automotive MES providers to verify their compliance with the current industry standards i.e. ISA-95 and ISA-88 using \textit{model-based requirement engineering} techniques. Moreover, as these standards are the building blocks for RAMI 4.0 verifying the compliance of MES products with them will act as a stepping stones towards the vision of Industry 4.0. The requirements model created for modeling the ISA-95 and ISA-88 needs, were made in SysML, which is a general purpose but standard language and allows to model a rich variety of relationships between different requirements. Use of SysML guarantees that these requirements will be easily understood without ambiguities by almost any vendor. The SysML based requirement models enables traceability analysis i.e., in case any requirements get updated or removed, then the tools that analyzes these models will help in detecting such modifications. Thus, one of the main reason why we choose to use model-based requirement modeling was to achieve consistency in our work and foster safety and quality of the software as its design models are evaluated in a precise manner~\cite{luo2016metrics,dajsurenevaluating}. Additionally, using SysML makes sure that any manufacturing stakeholder can create, use and extend these requirement models for more specific manufacturing sub-domains (in context of automotive domain) or update these models in a consistent manner as and when new standards for Industry 4.0 are available.

Additionally, during the vendor selection process, we choose the top four MES products wherein we choose one non-automotive specific MES tool to create the guidelines for benchmarking MES common features as this tool was fully compliant to ISA-95(100\%). In the course of our work, we made use of various sources such as whitepapers, blogs and articles. This was performed in order to avoid the bias (if any) in the industrial survey report~\cite{CGI_Group_MES_Product_Survey} that we used as the bases of our research for finding the MES vendors and their compliance to ISA-95. 

\section{Conclusions and Future Work}
\label{sec:conclusion}
In this paper, we provide an approach for bridging the gap between the current automotive MES tools landscape and the industry standards. We follow the model-based requirements engineering technique for finding, modeling and relating the needs to develop a state of the art MES tool using the ISA95 standards (building block for RAMI 4.0). We then perform a gap analysis between the modeled needs and the available MES tools. Thus, our work is mainly divided into three phases, (i) automotive MES tool selection phase, wherein we selected the MES products on the bases of their compliance with ISA-95 standards (ii) requirements modeling phase, wherein we model the requirements that we extracted from ISA-95 standard in SysML (iii) and gap analysis phase to illustrate the gap in compliance of the MES tools to the standard using the already modeled requirements (SysML models).

As a future work, we consider reusing and extending these requirement models for performing gap analysis for other domains of manufacturing. We aim to reuse these models for Industry 4.0 by engaging experts in a collaborative manner such as proposed by crowd-based requirement engineering~\cite{adepetu2012crowdrequire,groen2015towards}. This type of collaboration will be impossible without using standard-based requirement modeling techniques due to the ambiguity that may arise while using natural language text to express the needs or requirements in Industry 4.0.

\bibliographystyle{IEEEtran}
\bibliography{references}

\begin{thebibliography}{10}
\providecommand{\url}[1]{#1}
\csname url@samestyle\endcsname
\providecommand{\newblock}{\relax}
\providecommand{\bibinfo}[2]{#2}
\providecommand{\BIBentrySTDinterwordspacing}{\spaceskip=0pt\relax}
\providecommand{\BIBentryALTinterwordstretchfactor}{4}
\providecommand{\BIBentryALTinterwordspacing}{\spaceskip=\fontdimen2\font plus
\BIBentryALTinterwordstretchfactor\fontdimen3\font minus
  \fontdimen4\font\relax}
\providecommand{\BIBforeignlanguage}[2]{{%
\expandafter\ifx\csname l@#1\endcsname\relax
\typeout{** WARNING: IEEEtran.bst: No hyphenation pattern has been}%
\typeout{** loaded for the language `#1'. Using the pattern for}%
\typeout{** the default language instead.}%
\else
\language=\csname l@#1\endcsname
\fi
#2}}
\providecommand{\BIBdecl}{\relax}
\BIBdecl

\bibitem{westkamper2014manufacturing}
E.~Westk{\"a}mper, ``Manufacturing the backbone of the european economy,'' in
  \emph{Towards the Re-Industrialization of Europe}.\hskip 1em plus 0.5em minus
  0.4em\relax Springer, 2014, pp. 7--16.

\bibitem{hermann2016design}
M.~Hermann, T.~Pentek, and B.~Otto, ``Design principles for industrie 4.0
  scenarios,'' in \emph{2016 49th Hawaii International Conference on System
  Sciences (HICSS)}.\hskip 1em plus 0.5em minus 0.4em\relax IEEE, 2016, pp.
  3928--3937.

\bibitem{lu2016current}
Y.~Lu, K.~Morris, and S.~Frechette, ``Current standards landscape for smart
  manufacturing systems,'' \emph{National Institute of Standards and
  Technology, NISTIR}, vol. 8107, 2016.

\bibitem{hutchinson2014model}
J.~Hutchinson, J.~Whittle, and M.~Rouncefield, ``Model-driven engineering
  practices in industry: Social, organizational and managerial factors that
  lead to success or failure,'' \emph{Science of Computer Programming},
  vol.~89, pp. 144--161, 2014.

\bibitem{alferez2008model}
M.~Alf{\'e}rez, U.~Kulesza, A.~Sousa, J.~P. Santos, A.~Moreira, J.~Ara{\'u}jo,
  and V.~Amaral, ``A model-driven approach for software product lines
  requirements engineering.'' in \emph{SEKE}, 2008, pp. 779--784.

\bibitem{Rashid:2003:MCA:643603.643605}
\BIBentryALTinterwordspacing
A.~Rashid, A.~Moreira, and J.~Ara\'{u}jo, ``Modularisation and composition of
  aspectual requirements,'' in \emph{Proceedings of the 2Nd International
  Conference on Aspect-oriented Software Development}, ser. AOSD '03.\hskip 1em
  plus 0.5em minus 0.4em\relax New York, NY, USA: ACM, 2003, pp. 11--20.
  [Online]. Available: \url{http://doi.acm.org/10.1145/643603.643605}
\BIBentrySTDinterwordspacing

\bibitem{bombonatti2016usability}
D.~Bombonatti, C.~Gralha, A.~Moreira, J.~Ara{\'u}jo, and M.~Goul{\~a}o,
  ``Usability of requirements techniques: a systematic literature review,'' in
  \emph{Proceedings of the 31st Annual ACM Symposium on Applied
  Computing}.\hskip 1em plus 0.5em minus 0.4em\relax ACM, 2016, pp. 1270--1275.

\bibitem{isa95}
{ISA-95.00.03 Standard}, ``{ANSI/ISA-95.00.03-2013. Enterprise-control system
  integration - Part 3: Activity models of manufacturing operations
  management},'' International Society of Automation, 2013.

\bibitem{isa954}
{ISA-95.00.04 Standard}, ``{ANSI/ISA-95.00.04-2012. Enterprise-control system
  integration - Part 4: Objects and attributes for manufacturing operations
  management integration},'' International Society of Automation, 2012.

\bibitem{hankel2015reference}
M.~Hankel and B.~Rexroth, ``The reference architectural model industrie 4.0
  (rami 4.0),'' \emph{ZVEI}, 2015.

\bibitem{CGI_Group_MES_Product_Survey}
{CGI Group Inc.}, ``{Annual MES Product Survey Report},''
  \url{http://www.mescc.com/mes-report.html}.

\bibitem{Business_Software_advice_report}
Business-software.com, ``{Top 15 Manufacturing Software},''
  \url{http://www.business-software.com/offer/top-15-manufacturing-software/}.

\bibitem{Capterra_advice_report}
Capterra, ``{Top MRP Software Products},''
  \url{http://www.capterra.com/mrp-software/}.

\bibitem{software_advice_report}
{Software Advice}, ``{Compare Manufacturing Software},''
  \url{http://www.softwareadvice.com/manufacturing}.

\bibitem{plex_manufacturing}
{Plex System}, ``{Plex Manufacturing},'' \url{http://www.plex.com/}.

\bibitem{CTS_guide_manuf}
{CTS Guides}, ``{CTS Guide for Manufacturing},''
  \url{http://www.ctsguides.com/manufacturing/}.

\bibitem{CGI}
CGI, ``{MES Product Survey},'' \emph{MESA International}, vol.~16, no.~1, pp.
  1--768, 2015.

\bibitem{luo2016metrics}
Y.~Luo and M.~van~den Brand, ``Metrics design for safety assessment,''
  \emph{Information and Software Technology}, vol.~73, pp. 151--163, 2016.

\bibitem{dajsurenevaluating}
Y.~Dajsuren, ``Evaluating quality of automotive architectural and design
  models.''

\bibitem{adepetu2012crowdrequire}
A.~Adepetu, K.~A. Ahmed, Y.~Al~Abd, A.~Al~Zaabi, and D.~Svetinovic,
  ``Crowdrequire: A requirements engineering crowdsourcing platform.'' in
  \emph{AAAI Spring Symposium: Wisdom of the Crowd}, 2012.

\bibitem{groen2015towards}
E.~C. Groen, J.~Doerr, and S.~Adam, ``Towards crowd-based requirements
  engineering a research preview,'' in \emph{International Working Conference
  on Requirements Engineering: Foundation for Software Quality}.\hskip 1em plus
  0.5em minus 0.4em\relax Springer, 2015, pp. 247--253.

\end{thebibliography}

\end{document}